\documentclass[useAMS]{mn2e}
\title[Novae and second generation stars]{Novae from isolated white
dwarfs as a source of helium for second generation stars in globular
clusters} \author[Maccarone \& Zurek]{Thomas J. Maccarone\\ School of
Physics and Astronomy, University of Southampton, Hampshire SO17
1BJ,United Kingdom\\ \newauthor David R. Zurek\\ Department of
Astrophysics, American Museum of Natural History, New York, New York,
USA\\School of Physics and Astronomy, University of Southampton,
Hampshire SO17 1BJ,United Kingdom\\}

\begin{document}
\def\ltsim{\mathrel{\rlap{\lower 3pt\hbox{$\sim$}}
        \raise 2.0pt\hbox{$<$}}}
\def\gtsim{\mathrel{\rlap{\lower 3pt\hbox{$\sim$}}
        \raise 2.0pt\hbox{$>$}}}

\date{}

\pagerange{\pageref{firstpage}--\pageref{lastpage}} \pubyear{}

\maketitle

\label{firstpage}

\begin{abstract}
We explore the possible contribution of classical and recurrent novae
from isolated white dwarfs accreting from the intracluster medium to the
abundances of ``second generation'' globular cluster stellar
populations.  We show that under reasonable assumptions the helium
abundances of clusters can be enhanced substantially by these novae
and argue that novae should be considered as an important, and perhaps
even dominant channel in the evolution of the intracluster medium.  We
also discuss a possible test for whether helium enhancement really is
the cause of the multiple main sequences in globular clusters that is
independent of the positions of stars in the color-magnitude diagram.

\end{abstract}


\section{Introduction}
Globular clusters have traditionally been taken to be almost perfect
laboratories for studying the evolution of old stellar populations.
Until recently, it was thought that globular clusters were truly
simple stellar populations -- made of stars of uniform age and
chemical composition.  Over the past decade, it has become
increasingly clear that the stellar populations in globular clusters
are more complicated, with clear evidence for multiple main sequences
(Bedin et al. 2004), and giant branches; evidence for variations,
correlations and anticorrelations among light metals (e.g. Norris
1981; Gratton et al. 1986; Ivans et al. 1999; Marino et al. 2008); and
unusual horizontal branch (D'Antona et al. 2005) and subgiant branch
(Milone et al. 2008; Piotto 2009; Marino et al. 2009; Moretti et
al. 2009) morphologies that can be explained in straightforward ways
by the presence of multiple stellar populations.

The apparent second generation populations in the globular clusters
tend to share a set of chracteristics that are difficult to explain in
terms of normal chemical evolution.  In particular, the fact that the
metal rich population in $\omega$ Cen is bluer, not redder, than the
metal poor population, argues for strongly enhanced helium abundances
in the younger populations (e.g. Norris 2004; Piotto et al. 2005).  In
most other clusters, the age difference between the older and younger
population seems to be no more than a few hundred million years and
the metallicities appear to be very similar (e.g. Piotto 2009).  In
fact, enhanced helium abundances can explain nearly all the
phenomenology of multiple stellar populations seen in globular
clusters (e.g. D'Antona et al. 2011), and, largely speaking, most of
the debate in recent years has been about how to produce such a large
helium abundance, rather than whether enhanced helium abundances are
really the proper solution to the problem, although there have still
been suggestions that clusters with multiple stellar populations might
have formed from mergers of two clusters (e.g. Mackey \& Broby Neilsen
2007).

Two channels have been proposed and debated in recent years for
producing large helium abundances in globular clusters' second
generation stars.  One is enrichment from the winds of fast-rotating
massive stars (e.g. Prantzos \& Charbonnel 2006; Decressin et
al. 2007).  The other is the enrichment from the ejecta of AGB stars
(first suggested by Cottrell \& DaCosta 1981 in the context of sodium
and cyanogen anomalies, and applied to the helium abundance problem by
e.g. D'Antona \& Ventura 2007; D'Ercole et al. 2008, 2010)

On the other hand, the most detailed model papers published to date
invoke very unusual initial mass functions for the clusters in order
to allow the production of a large enough amount of helium from a
small enough number of first generation stars (e.g. truncation at 9
$M_\odot$ for the second generation -- D'Ercole et al. 2010; or an
initial mass function for the first generation considerably steeper
than a Kroupa IMF -- Prantzos \& Charbonnel 2006).  In some cases,
substantial stripping of the stars on the outer part of the cluster is
needed to produce enough helium enrichment in the second generation of
stars, while also having a large enough ratio of the number of low
mass second generation to first generation stars in the present epoch
(e.g. D'Ercole et al. 2008).  While none of these issues is, on
observational grounds problematic (e.g. it is very easy to hide many
first generation cluster stars in the Galactic halo, and the
combination of stellar and dynamical evolution in clusters makes it
difficult to extract their initial mass functions from their present day
mass functions), the need for a series of potentially uncomfortable
assumptions may indicate that the current theory does not tell the
whole story of helium enhancement in globular clusters.  As a result,
alternative mechanisms for helium enrichment which might operate
either instead of those already proposed, or in conjunction with them,
are well worth consideration.

In this Letter, we discuss a new scenario for enhancing the helium
abundance of the gas that forms the second generations in globular
clusters: novae powered by accretion of the intracluster medium by
massive white dwarfs.  Given the uncertainties in both the nova
yields, and the likely properties of the clusters early in their
lifetimes, we present only toy calculations at this stage to show that
this mechanism is capable of producing a substantial yield of helium,
rather than detailed calculations aimed at reproducing observed
clusters.  First we outline our assumptions about the rate of
accretion of material from the interstellar medium.  Next we show what
the rates of helium production would be under those assumptions, based
on existing estimates of nova yields.  Additionally, we show that the
helium-enchancement hypothesis is potentially testable by looking for
mass segregation in clusters, since the helium rich stars should be
lighter than the helium poor stars at the same luminosity.

\section{The amount of accretion from the ISM}
To date, there are no stellar mass compact objects which show clear
evidence for strong accretion from the interstellar medium.  A few
isolated neutron stars represent candidates for accretion from the
ISM.  They are most likely relatively young neutron stars still
cooling from their initial formation (e.g. Perna et al. 2003; Kaplan
et al. 2011), although in at least one case, there is evidence for a
hydrogen atmosphere on an isolated neutron star, which can be
explained as evidence for accretion, but could also be due to
diffusive nuclear burning (e.g. Ho et al. 2007; Chang et al. 2004).

On the other hand, the Bondi accretion rate is expected to be highest
for objects located deep within molecular clouds or in other regions
where the interstellar medium is especially dense -- precisely the
regions where foreground absorption and scattering of the accretion
light would make sources hardest to observe.  It has been suggested
that such sources might be most easily detected in the radio,
especially with the upcoming generation of wide field, low frequency
radio arrays (Maccarone 2005), but at the present time adequate
searches have not been made.  The possibility that the nova rate in
star clusters might be enhanced by accretion from intracluster gas has
been discussed previously by Naiman et al. (2011).

It is clear that the site of formation of second generation
populations in globular clusters must be a region of gas density
comparable to that in molecular clouds (since the total amount of star
formation must be of order $10^5-10^6 M_\odot$, and the size scale for
the region must be of order 1 pc).  Assuming one starts with $10^5
M_\odot$ within 1 pc of the cluster center, the gas density will be
about $10^6$ particles per cubic centimetre.  

We use the results of Dobbie et al. (2006) to estimate the birth
masses of the white dwarfs present in the cluster:
\begin{equation}
M_{WD}=0.7+(M_i-3.0)/8
\end{equation}
where $M_{WD}$ is the initial mass of the white dwarf, and $M_i$ is
the initial mass of the star, both in solar units.  We can next
estimate the fraction of the initial stellar population that will be
in white dwarfs produced by progenitors of at least $3 M_\odot$. By
integrating over a Kroupa IMF from 0.08 to 80 solar masses, we can get
the total initial mass of the cluster.  Then, by integrating the mass
of the white dwarfs using the Dobbie et al (2006) relation, we find
that about 8\% of the initial mass of the first generation ends up in
white dwarfs.  We also find that about 50\% of the initial mass of the
first generation is in main sequence stars that still exist.

Rescaling the Bondi-Hoyle rate formula from Ho, Terashima \& Okajima
(2003), we find that:
\begin{equation}
\dot{M}_b=7\times10^{-9} M_\odot {\rm yr}^{-1} \left(\frac{M_{WD}}{M_\odot}\right)^2 \left(\frac{n}{10^6 \rm{cm^{-3}}}\right) \left(\frac{c}{10^6 \rm{cm/sec}}\right)^{-3}
\end{equation}
which gives mass accretion rates of order $10^{-8} M_\odot {\rm
yr}^{-1}$ over most of the cluster core.  We compare the parameter
values above with those in the figures in D'Ercole et al. (2008), who
estimate that the central gas density should be $\sim10^6$ cm$^{-3}$,
with central temperatures of a few thousand K, yielding slightly
higher expected mass accretion rates than $10^{-8}$ $M_\odot$
yr$^{-1}$ in the very centres of the clusters.  They find that the gas
density falls off a bit toward the outer part of the cluster core, but
that the temperature does as well.

We then estimate how much accretion could take place over the
$\sim300$Myr during which white dwarfs exist, but before the second
generation of stars has formed.  One can see that it is not
unreasonable for $\sim3 M_\odot$ of gas to be processed by each solar
mass of white dwarf -- although only the heaviest white dwarfs will
have existed for the full 300 Myrs available, so it is more likely that
the average white dwarf will process roughly its own mass rather than
three times its own mass.  According to the calculations of Yaron et
al. (2005), for mass accretion rates of $10^{-8} M_\odot {\rm
yr}^{-1}$, the heaviest white dwarfs will be very near balance between
the mass accreted and the mass lost in the nova shells, so the masses
of the white dwarfs themselves can be assumed not to change
substantially.  This will remain true for white dwarf masses greater
than about 0.8$M\odot$, which represent most of the white dwarf mass
that will exist in a cluster with a Kroupa (2001) IMF over the allowed
timespan.

We do note that in some cases, there is evidence that the Bondi
formula significantly overestimates the mass accretion rates of
objects accreting from the interstellar medium.  Perna et al. (2003)
find that the deficit of isolated neutron stars accreting from the
interstellar medium can be explained by suppressing accretion from the
interstellar medium due to magnetic pressure effects in the ISM
(e.g. Igumenchev \& Narayan 2002), but also noted that propeller
effects from the neutron stars' own magnetic fields could be
responsible for keeping the neutron stars from accreting
substantially.  Pellegrini (2005) shows that active galactic nuclei in
elliptical galaxies are typically accreting at about 3\% of the Bondi
rate -- but a variety of processes may act to suppress accretion in
galactic nuclei, included e.g. radiative and kinetic feedback from the
accretor itself (e.g. Milosavljevic et al. 2009).  On a more positive
note, in high mass X-ray binaries, the mass transfer rates are well
modelled by the Bondi capture process (see e.g. Frank, King \& Raine
2002).  Finally, as we show below, our simple calculation gives an
excess of helium compared to what is needed to match the observations
if we make assumptions similar to those in D'Ercole et al. (2008)
about the properties of the first generation and the gas reservoir
that supplies the second generation, so it is easily possible to
tolerate some mild suppression of the accretion rate below the Bondi
capture rate.

\section{The expected nova yields}
It is generally well-agreed that the properties of novae will depend
strongly on the white dwarf mass, white dwarf temperature, and the
mass transfer rate onto the white dwarf.  The yields of helium and
metals, and the ejected masses seem to depend only weakly on the
composition of the white dwarf, while the specific composition of the
metals in the ejecta can depend strongly on whether the white dwarf is
a carbon-oxygen white dwarf or a neon-oxygen-magnesium white dwarf
(see e.g. Yaron et al. 2005).  The set of parameter values for which
nova yields have been calculated has been steadily expanding over the
past decade, but is still not detailed enough to account for details
of the metal abundances of gas produced from novae (e.g. the
calculation of Yaron et al. 2005, which is probably the most
sophisticated treatment to date, treats ONeMg white dwarfs with a
calculation assuming the white dwarf is only made of oxygen and neon).
Similarly, nearly all calculations done to date assume that the
accreted material is of solar composition (although see also Jos\'e et
al. 2007, who do a calculation of the yields from very low metallicity
gas as might be found in Population III).

Next, we consider the possibility that the nova are enriching largely
pristine gas.  This assumption allows for an arbitarily large gas mass
to be used to form the second generation of stars relative to the mass
of the first generation.  It also requires that the gas not be unbound
entirely from the cluster by the core collapse supernovae from the
first generation, and also that the gas not be too strongly enriched
by the core collapse supernovae.  Scenarios have been developed
already in which the pristine gas can survive in this manner
(e.g. Recchi et al. 2001).  In this manner, if the gas can be
sufficiently enriched by novae, then the second generation can have a
similar stellar mass to that of the first generation if (1) only a
small fraction of the mass in the first generation is turned into
stars and (2) most of the gas is blown from the core of the cluster by
the first generation's core collapse supernovae, but is not actually
unbound from the cluster, so that the second generation is formed from
a total gas mass similar to that from which the first generation is
formed and (3) the star formation efficiency (i.e. the stellar mass
produced per unit gas mass) for the second generation is similar to
that for the first generation.  In this way, nova enrichment might
solve the problems posed by having the helium rich material be a small
fraction of the gas released from first generation stars (either
through equatorial or AGB winds), but being a large fraction of the
total mass of the first generation stars which remain.  The existent
models (D'Ercole et al. 2008; Decressin et al. 2007) require a large
fraction of the first generation's stars to be stripped from the
outsides of the clusters -- a process which is plausible, but perhaps
not as attractive as a model in which the mass evolution required of
the clusters is not so extreme.

We take the yields from the calculations of Yaron et al. (2005) as
instructive, but not as definitive.  The helium content of the novae
ejecta are typically about 0.4-0.5 for high accretion rates.  We then
expect the final helium content of the gas to be about
$0.21\times$$M_{WD}/M_{gas,sec}+0.24$, where $M_{WD}$ is the mass
locked up in the white dwarfs from the first generation of star
formation, and $M_{gas,sec}$ is the mass of the gas that eventually
produces the second generation.  To reach $Y=0.30$, we then need to
have the mass processed in novae to be about (0.06/0.21) of the mass
in the gas that will make up the second generation.  This level is
comfortably achieved in, e.g., the modeling currently used to produce
second generations (e.g. D'Ercole et al. 2008).  The white dwarfs
represent about 8\% of the mass of the first generation by the time
the 3 $M_\odot$ main sequence stars form white dwarfs.  Since a white
dwarf can process roughly its own mass, and the first generation is a
few to ten times as massive as the second generation, it would be
surprising if novae did not significantly enhance the helium content
of clusters -- in fact, the amount of helium we would naively expect
to be produced in novae is so large that the fact that clusters are
not routinely observed with $Y$=0.4 suggests that perhaps the Bondi
capture rate is an overestimate for white dwarfs accreting from the
interstellar medium.  By providing an extra channel for helium
enhancement, then, one can relax the assumptions about the second
generation's initial mass function.  Evaluating the effects of novae
on the abundances of the light metals like oxygen, aluminium, sodium,
carbon and nitrogen would require yield calculations far more complex
than the ones which have been made to date.  However, given the
suggestions that some of the rarer isotopes of carbon, nitrogen and
oxygen (i.e. $^{13}$C, $^{15}$N and $^{17}$O) may be produced
primarily in novae (e.g. Romano \& Matteuci 2003), it may be
worthwhile to check whether the isotopic abundances of these elements
in red giants in globular clusters with multiple populations show
anomalies.

We note that gas must be cycled through multiple nova explosions
before it becomes fully enriched -- the ejecta of a single nova may
reach a helium abundance of $\sim0.3-0.4$, but the ejecta will travel
over a distance of $10^{-3}$ pc (see section 4) before the nova shell
stops expanding.  As a result, the nova shell will sweep up many times
its own mass, and the gas will not be sufficiently enriched by a
single nova to form second generation stars.  

The nova yields of metals are particularly uncertain theoretically,
but are likely to be keys to understanding whether novae are important
contributors to the chemical enrichment of globular clusters' second
generation stars.  The relevant novae are likely to be novae from
ONeMg white dwarfs.  One can reach this conclusion through two lines
of reasoning -- first the ONeMg white dwarfs are the heaviest and form
first, so they will have the bulk of accretion take place onto them
before the second generation of stars form (since the second
generation must form fairly rapidly).  Secondly, they produce the
largest yields of helium, and, in some accretion rate regimes, can
account very nicely for a strong sodium-oxygen anticorrelation (Yaron
et al. 2005).  We note that the progenitors of these white dwarfs will
predominantly be the same stars that are often taken to supply helium
enriched material to the interstellar medium (D'Ercole et al. 2008).
As a result, even if a susbtantial reservoir of pristine gas provides
the bulk of material for accretion by the white dwarfs, its
composition will be affected at least in part by the addition of the
helium rich AGB ejecta that are released as the white dwarfs are
forming.

At the present time, only a few observational data sets have been
obtained on ONeMg abundances (e.g. Schwarz et al. 2007).  Generically,
these novae produce very large enhancements of nitrogen and neon
relative to solar composition, and frequently, they produce large
magnesium enchancements as well.  They also typically lead to a helium
mass fraction of about 32\%, although with significant scatter.  A
potential problem is that mild oxygen enhancement is seen in most
cases, while oxygen depletion is observed for the candidate
helium-rich stars in globular clusters (e.g. Marino et al. 2011).

The models of Yaron et al. (2005) {\do} predict oxygen depletion and
sodium enhancement for most ONe novae.  Whether the discrepancy exists
because of problems with the theoretical nova models or observational
errors or selection biases in the small number of ONeMg novae which
have been observed is difficult to determine at the present time.
However, if a large number of high accretion rate novae from massive
ONeMg white dwarfs are seen to produce excesses of oxygen relative to
solar values, this should be taken as a point against the nova
enrichment scenario -- although there will still remain the point that
the chemical composition of the gas in globular clusters early in
their lifetimes will be rather different than the gas accreted for
most present-day novae.

\section{Stability of the cluster against novae clearing out the gas}
It has recently been suggested by Moore \& Bildsten (2011) that most
globular clusters today may have low gas content because novae
frequently sweep out all the gas from the cluster.  The clusters we
are considering should be robust to having even the high rate of novae
we are considering here from sweeping out their gas.  The momentum
carried by a shell of $10^{-5} M_\odot$ at about 1000 km/sec (typical
numbers for a 1.2 $M_\odot$ WD accreting at about $10^{-8} M_\odot
{\rm yr}^{-1}$) should be large enough to evacuate a bubble of radius
about $10^{-3}$ pc in gas with random motions on the order of 10
km/sec, before the outflow speed is comparable to the random motions
of the gas.  The re-filling timescale of the bubble would be of order
30-100 years.  As a result, the novae would provide some energy input
into the cluster, but would not blow gas out of the cluster.  This is
unsurprising, since the D'Ercole et al. (2008) have already pointed
out that a single Type Ia supernova would represent only a mild
perturbation on the cluster's gas reservoir, and the sum of energy
injections from all the novae would be significantly less than that
from a single supernova -- although they also went on to point out
that a large rate of supernovae would have severe effects on the gas
reservoir of the cluster.

One can alternatively look at the filling factor of the bubbles that
are expected to be created by nova explosions.  With $\sim10^5$ white
dwarfs in the cluster, each undergoing a nova explosion every
$\sim1000$ years, there should be about 100 novae per year in the
cluster.  Since the lifetimes of the nova bubbles are $\sim100$ years,
there should be about $10^4$ active bubbles at a time, each of which
has a radius of $\sim10^{-3}$ pc, and hence a volume of about
$10^{-8}$ pc$^3$.  Therefore, $\sim10^{-4}$ of the cluster core volume
should be filled by the bubbles, and they can be neglected in terms of
the effects of interactions of bubbles with one another, and the fact
that the accretion rates within a bubble will be much lower than those
outside a bubble.

\section{Effects of accretion on the first generation stars}

One can also consider whether the first generation's stars can be
expected to be affected significantly by accretion of interstellar
gas.  In considering this, it is important to bear in mind two points:
first, in most clusters, nearly all the stars below the turnoff mass
at the present time are fully convective.  Furthermore, none is more
than about half the mass of the heavy white dwarfs from which the
recurrent novae that are expected to pollute the interstellar medium
are produced.  The former effect means that whatever gas has been
accreted by a main sequence star in a globular cluster should have
been well-mixed by the present time (see e.g. the discussion in Briley
et al. 2002 of why accretion is unlikely to lead to the unusual
abundances of some M~13 subgiants).  One can then consider what
fraction of the mass of the present day main sequence stars would have
come from accretion.  Integrating equation 1, we find that in the
inner parts of the cluster, $M_f-M_i = 2 M_f M_i$, where $M_f$ is the
final mass of the star, and $M_i$ is the initial mass of the star,
with both quantities divided by the solar mass.  Over most of the
first generation's main sequence, stars should have acquired about
half their masses by accretion, much of which will have taken place
before much enrichment of the gas, so the devations of their
abundances from those of stars which formed from first generation
material and did not accrete at all should be only about a quarter as
big as the differences between the first and second generation stars,
meaning that these stars should probably be hard to identify with
current data quality.

\section{What novae cannot explain}
It is also important to consider what observed phenomena cannot be
explained by nucleosynthesis taking place in classical novae.  Largely
speaking, classical novae do not burn beyond chlorine.  Therefore,
some of the heavier s-process elements recently shown to have unusual
abundances (e.g. NGC~6656 -- Milone et al. 2011) which cannot be
produced by novae.  On the other hand, Milone et al. (2011) suggest
that some supernova enrichemnt was probably likely for the second
generation in NGC~6656, as some of the stars also show enrichment in
elements such as europium, which are not thought to be produced except
in supernovae -- but of course these elements represent a challenge
also to the scenarios in which the enrichment comes from stellar
winds.  Additionally, we again emphasize the possibility that novae
may act in concert with other mechanisms to change the composition of
the gas from which second generation stars form.

Another example of something which is difficult to explain in the
scenario we propose is the observation of a strong sodium-oxygen
anticorrelation in M3 (Cohen \& Mel\'endez 2005), with no evidence for
a helium enhancement (Catelan et al. 2009).  Because the results of
Cohen \& Mel\'endez (2005) extended the discovery of the sodium-oxygen
anticorrelation down to masses on the red giant branch below where
dredge-up should have taken place, they argue that some form of
pollution must be necessary for these systems, and the lack of such
correlations for field stars (Gratton et al. 2000) suggests that the
origin of the pollution must be inherently related to the dense
stellar environments, but attempts to model such pollution have not,
to date, been successful (e.g. Fenner et al. 2004).  These
observations remain a problem in the context of all current models of
the chemical evolution of globular cluster stars.

\section{Mass segregation as a test of the helium hypothesis}

We also discuss a possible test for the helium enrichment hypothesis
for producing the multiple main sequences and the other anomalies seen
in evolved stars in globular clusters.  No direct measurements of the
helium contents of stars suspected to be helium enhanced have been
possible to date, leaving the helium enhancement scenario largely
untested, except based on its effects in color-magnitude diagrams.  An
alternative indirect test for the helium enrichment scenario is that
the masses of stars at a given luminosity should be quite different.
For example, D'Antona \& Caloi (2004) calculate three tracks with the
metallicity and age of NGC~2808, but with helium abundances of 0.24,
0.28 and 0.32.  They find that the turnoff luminosities for the three
tracks are the same, but that the turnoff masses are 0.82, 0.77 and
0.72, respectively.  It should be expected, then, that the core radius
for the helium rich stars will be larger than the core radius for the
stars with normal helium abundances due to mass segregation (see
e.g. Gunn \& Griffin 1979, since all globular clusters in the Galaxy
are older than their core relaxation timescales (Harris 1996 and
references within).  On the other hand, a few clusters do have
relaxation timescales at their half-light radii which are longer than
their ages (Harris 1996 and references within), so searching for mass
segregation of the second generation well outside the core could lead
to spurious results, especially since most models predict that the
second generation formed in a more centrally concentrated manner than
the first generation.

We note that because the mass functions may differ for the different
generations, a straight test of the spatial profile of the red
sequence versus that of the blue sequence could potentially be
misleading.  Additionally, because the clusters may not be fully
relaxed at radii observable from the ground, ground-based data is
unlikely to be useful.  We have tested the feasibility of this
mechanism by drawing stars at random from a King (1966) model
distribution.  We take a King model with $W=7$ (corresponding to a
ratio of tidal to core radius of about 33 or a concentration parameter
in the units of the Harris catalog of about 1.5), and repeatedly draw
about 1400 stars at random (because we draw the stars in each
spherical shell as a Poisson distribution, the number varies a bit
from simulation to simulation).  Typically about 100 of those stars
have projected radii less than the core radius of the cluster.  We
then compute the mean value for the distance of these stars from the
cluster centre, and find that it is about 0.48$\pm0.02$.  Assuming
that the stars really fit to a Gunn-Griffin (1979) model, groups of
100 stars in the red and blue sequences should then yield $\approx$
3$\sigma$ differences in terms of their mean distances from the centre
of the cluster if they have masses of 0.82 and 0.72 $M_\odot$,
respectively.  Many such bins can be constructed in the clusters
massive enough to show multiple main sequences, in a manner that keeps
the luminosity bins having small mass ranges within themselves, and
then one can look to see whether a consistent trend holds up.
Previous approaches have typically looked at the spatial profiles of
the different main sequences as a whole, rather than breaking the
stars down into luminosity bins (e.g. Sollima et al. 2007; D'Ercole et
al. 2008).

At the present time, we are aware only of observations of the outer
part of $\omega$~Cen which have been used to compare the red and blue
main sequences (Sollima et al. 2007).  $\omega$~Cen shows that the
blue (i.e. likely helium rich) main sequence is centrally
concentrated.  On the other hand, the observations compare the spatial
profiles of the two main sequences on physical scales with relaxation
timescales of order or greater than a Hubble time (the half-light
radius relaxation time for $\omega$~Cen is about 12 Gyrs -- Harris
1996), so one would not expect the mass segregation effects to have
taken place yet.

\section{Discussion}

We have shown that the production of a substantial amount of helium
from novae triggered by accretion of intracluster medium by isolated
white dwarfs is a necessary consequence of having a large amount of
gas available in a globular cluster's core well after the white dwarfs
have started forming.  We have also outlined a method that can be used
for testing whether the second generation populations of globular
clusters really are helium enhanced.

\section{Acknowledgments}
We thank Mike Shara, Christian Knigge, Brian Warner and Grace Thomson
for useful discussions on topics directly related to this paper, and
Kelly Holley-Bockelman for useful discussions several years ago on a
different topic related to gas in the early epochs of globular
clusters.  We thank the referee Raffaelle Gratton for constructive
criticism which substantially improved the paper.

\label{lastpage}

\end{document}